# A sliding mode multimodel control for a sensorless photovoltaic system

Ahmed Rhif, Zohra Kardous and Naceur BenHadj Braiek
Advanced System Laboratory, Polytechnic School of Tunisia (Laboratoire des Systèmes Avancés(LSA), Ecole Polytechnique de Tunisie, BP 743, 2078 La Marsa, Tunisie

This study presents a sliding mode multimodal control for a sensorless photovoltaic panel, which sets itself as per the sun position at every second during the day for a period of five years; then tracker, using sliding mode multimodal controller and a sliding mode observer, will track these positions to make sunrays orthogonal to photovoltaic cell that produces more energy. After sunset, tracker goes back to initial position of sunrise. This autonomic dual axis sun tracker increases power production by over 40%.

**Keywords:** Altitude, Azimuth, Photovoltaic panel, Solar energy, Sun tracker

## Introduction

Photovoltaic arrays are fixed. In this way, solar energy incident on modules is not optimal during time variation and seasons. Theoretical studies[1-2] show that energy supplied by a photovoltaic cell equipped by a sun tracker (ST) is higher (30-40%) than energy provided by fixed panels (Fig. 1). A solar panel (3 m²) fixed to a surface produces 5 kWh of electricity per day. The same installation, if equipped with a tow axis sun ST[3-5] (Fig. 2), can provide up to 8 kWh per day. This study presents speed and position control of ST using sliding mode control (SMC) and a sliding mode observer (SMO).

## Experimental Section
### Dual Axis Sun Tracker (ST) Modelling

A dual-axis tracking system with two degrees of freedom (DOF) was used for following the sun position in both altitude and azimuth directions, keeping the sun's rays normal to photovoltaic cell. Operational subset of ST has two DOF, motorized by tow stepper motors, in order to track exactly the prescribed sun's path. Mechanical system (Fig. 3) consists of altitude motor to drive horizontal axis, azimuth motor to drive vertical axis and photovoltaic cell (150 mm x 100 mm). It was designed using solidWorks software in the Department of Electronics, High Institute of Applied Sciences and Technologies Sousse (ISSAT so), and realised in Plexiglas profiles in order to obtain a simple and economic structure.

Considering that system operation is accomplished by one of those stepper motors at each time, mathematical model of the process in d-q (direct-quadrature) transformation[6] (Fig. 4) would be given as

$$\begin{cases} \dfrac{di_d}{dt} = \dfrac{1}{L}\left(v_d - Ri_d + NL\Omega\ i_q\right) \\ \dfrac{di_q}{dt} = \dfrac{1}{L}\left(v_q - Ri_q + NL\Omega\ i_d - K\Omega\right) \\ \dfrac{d\Omega}{dt} = \dfrac{1}{J}\left(Ki_q - f_v\Omega - C\right) \\ \dfrac{d\mathbf{q}}{dt} = \Omega \end{cases} \quad \ldots(1)$$

where $R$, resistance; $L$ inductance, $J$, inertia; $N$, spin number; $K$, torque constant gain; $f_v$, friction; $\Omega$, angular velocity; $\mathbf{q}$, rotor position; $i_d$ and $i_q$, currents in d-q referential; and $v_d$ and $v_q$, voltages in d-q referential.

Considering $\begin{cases} x = [i_d, i_q, \mathbf{q}, \Omega]^T \\ u = [v_d, v_q]^T \end{cases}$, then model state function will be given as

$$\dot{x} = f(\Omega)x + gu + p \quad \ldots(2)$$

---
*Author for correspondence
E-mail: ahmed.rhif@gmail.com

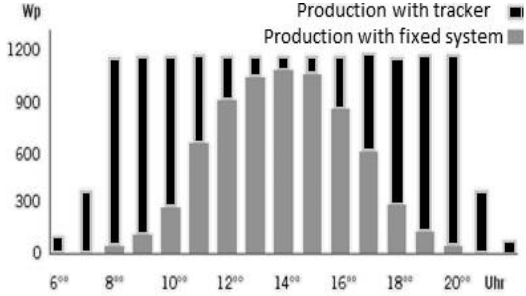

Fig.1—Histogram of solar energy produced

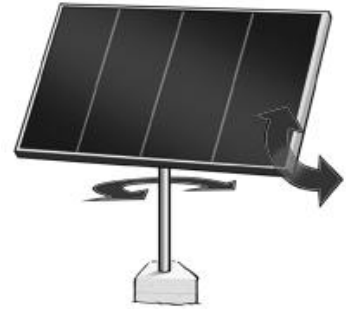

Fig.2—Two axis system orientation

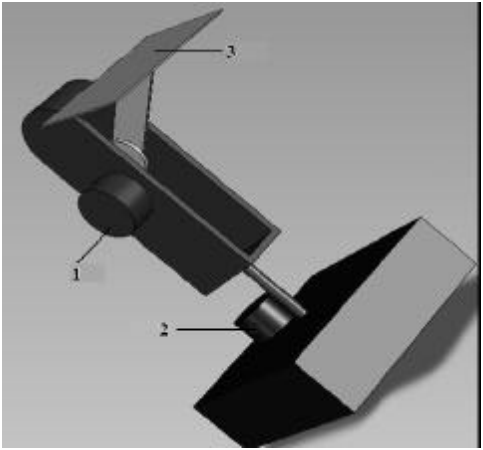

Fig. 3—Mechanical design of the process

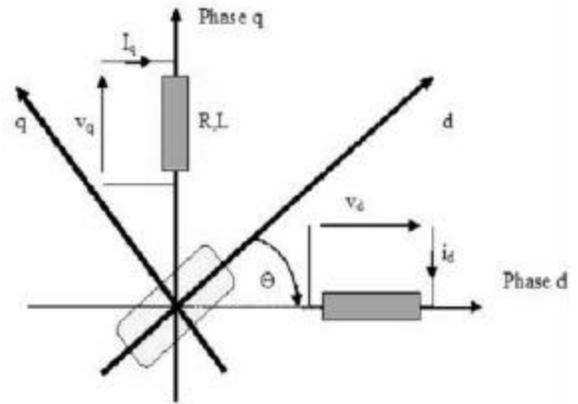

Fig.4—Electrical model in d-q referential

where

$$f(\Omega)=\begin{bmatrix}-\frac{R}{L} & N\Omega & 0 & 0 \\ N\Omega & -\frac{R}{L} & a_{23} & -K \\ 0 & 0 & 0 & 1 \\ 0 & \frac{K}{J} & 0 & -\frac{f_v}{J}\end{bmatrix} \quad g=\begin{bmatrix}\frac{1}{L} & 0 \\ 0 & \frac{1}{L} \\ 0 & 0 \\ 0 & 0\end{bmatrix} \quad p=\begin{bmatrix}0 \\ 0 \\ 0 \\ -\frac{C}{J}\end{bmatrix} \text{ with}$$

$f$ and $g$ as linear functions.

In other way, Eq. (2) could be given in a simplest way; considering $y_1 = q$ and $y_2 = i_d$, Eq. (1) gives

$$\begin{cases} q = y_1 \\ \Omega = \dot{y}_1 \\ i_d = y_2 \\ i_q = \frac{1}{K}(J\ddot{y}_1 + f_v\dot{y}_1 + C) \\ v_d = L\dot{y}_2 + Ry_2 - \frac{NL}{K}\dot{y}_1(J\ddot{y}_1 + f_v\dot{y}_1 + C) \\ v_q = \frac{JL}{K}y_1^{(3)} + \frac{1}{K}(Lf_v + RJ)\ddot{y}_1 + \left(\frac{Rf_v}{K} + K + NLy_2\right)\dot{y}_1 + \frac{L}{K}\dot{C} \end{cases} \quad \ldots(3)$$

**Controller Designs**

Among controller types tested and implemented in ST, Hua & Shen[1] compared sun tracking effectiveness of maximum power point tracking (MPPT) algorithms with a control method, which combined a discrete control and a Proportional Integral (PI) controller to track maximum power points of a solar array[7-10]. Yousef[11] developed a ST system with a fuzzy logic control for nonlinear dynamics of tracking mechanism[12,13]. Roth[14] designed a solar tracking system to measure direct solar radiation. The system was controlled by a closed loop servo system composed by four quadrant photo-detectors needed to sense the position of sun[15,16]. Finally, a sensorless control test using SMC, which is very solicited in tracking path studies.

*High Order Sliding Mode Control (SMC)*

SMC can bring back the system state on sliding surface, where it will slide along the desired state. However, SMC needs a high level of discontinuous control, which makes harmful effects on actuators, known as chattering phenomenon (CP). As a solution, designing of a high order SMC consists of sliding variable derivative computing[17-22]. A second order SMC consists of $s(x) = \dot{s}(x) = 0$ and $\dot{V} = \frac{1}{2}\frac{\partial}{\partial t}(s^2) = s\dot{s} \leq -h|s|$ with $h > 0$ and $V$ as Lyaponov quadratic function. Path tracking error in the origin is stabilized as $e = [i_d - i_{dr}, i_q - i_{qr}, \Omega - \Omega_r, q - q_r]^T = [e_1, e_2, e_3, e_4]^T$. Using Eq. (1), one gets

$$\begin{cases} \dot{e}_1 = \frac{1}{L}(v_d - v_{dr} - Re_1 + NL(e_3 e_2 + e_3 i_{qr} + e_2 \Omega_r)) \\ \dot{e}_2 = \frac{1}{L}(v_q - v_{qr} - Re_2 + NL(e_3 e_1 + e_3 i_{dr} + e_1 \Omega_r) - Ke_3) \\ \dot{e}_3 = \frac{1}{J}(Ke_2 - f_v e_3 - C_r) \\ \dot{e}_4 = e_3 \end{cases} \quad \ldots(4)$$

where $e$, system error; $i_{dr}$ & $i_{qr}$, reference current in d-q referential; $v_{dr}$ & $v_{qr}$, reference voltage in d-q referential, $\Omega_r$ reference angular velocity; and $q_r$, reference rotor position.

For velocity control by a second order SMC, sliding surface is written as $s_\Omega = me_3 + \dot{e}_3$ with $m > 0$. First order derivative of Eq. (4) gives $\dot{s}_\Omega = m\dot{e}_3 + \ddot{e}_3$ and $\dot{s}_\Omega = m\frac{1}{J}(Ke_2 - f_v e_3 - C_r) + \frac{1}{J}(K\dot{e}_2 - f_v \dot{e}_3 - \dot{C}_r)$. In second phase, to deal with position control process, sliding surface is considered as

$$s_q = m_1 e_4 + m_2 \dot{e}_4 + \ddot{e}_4 \quad \ldots(5)$$

Combining Eq. (4) and Eq. (5) gives $s_q = m_1 e_4 + m_2 e_3 + \frac{1}{J}(Ke_2 - f_v e_3 - C_r)$ with $m_1, m_2 > 0$. In convergence phase, to bring the system on sliding surface, different forms of switching control are available. In this study, the following were chosen ($U_0 > 0$): $u_{s1} = -U_0 sign(s_\Omega)$; and $u_{s2} = -U_0 sign(s_q) = \dot{s}_q$.

**Proposed Sliding Mode Multimodel Control (SM-MMC) Synthesis**

Multimodel approach represents an interesting alternative to SMC problems, and a powerful tool for identification, control and analysis of complex systems. Multimodel principle makes possible the design of a non linear control composed by linear sub controls relative to each sub model. The process control could be then deduced from a fusion or a commutation between different partial controls weighted by equivalent validities. In this sense, several methods of validities estimation [ $v_{in}(t) = \frac{v_i(t)}{\sum_{i=1}^{N} v_i(t)}$ and $v_i(t) = 1 - r_{in}(t)$ ] are already reported. These methods are classified according to the models acquisition ways, which are related to process knowledge. This paper will apt for residue approach for validities computing as

$$r_{in}(t) = \frac{r_i(t)}{\sum_{i=1}^{N} r_i(t)} \quad \ldots(6)$$

$$r_i(t) = |y(t) - y_i(t)| \; ; \; i = 1,\ldots,N \quad \ldots(7)$$

where $v_i$, validity; $r_i$, residue; $y(t)$, system's output; and $y_i(t)$, output of $i^{th}$ model.

In order to reduce disturbances due to inadequate models, validities were reinforced as

$$v_i^{renf}(t) = v_i(t) \prod_{\substack{j=1 \\ i \neq j}}^{N} (1 - v_j(t)) \quad \ldots(8)$$

Normalized reinforced validities are given as

$$v_{in}^{renf}(t) = \frac{v_i^{renf}(t)}{\sum_{i=1}^{N} v_i^{renf}(t)} \quad \ldots(9)$$

SM-MMC uses multimodel fusion approach to reduce switching control effect, which induces CP[23-24]. To improve control performances, Ure & Inalhan[25] applied a discrete sliding mode. In another study[26], a non linear control was considered for a combat air vehicle to execute agile manoeuvres. To reduce CP, fuzzy mode was applied to a high fidelity six degrees of freedom F-16 fighter aircraft model. In this study, control approach consisted in carrying out a fusion on sliding mode discontinuous control (Fig. 2) to eliminate or minimize CP. To adapt the controlling process to each sub model, several sliding surfaces were used, each state of a sub model $M_i$ was considered to reach one of these sliding surfaces $s_i$ (Fig. 5). To ensure SMC existence, several switching control $u_{si}$ relative to each sliding surface $s_i$ were used. Then, partial controls $y_i$ of each sub model will contribute on validities calculation as

$$u_{si} = \begin{cases} u_{si\,min} & si \quad sign(s) < 0 \\ u_{si\,max} & si \quad sign(s) > 0 \end{cases} \quad \ldots(10)$$

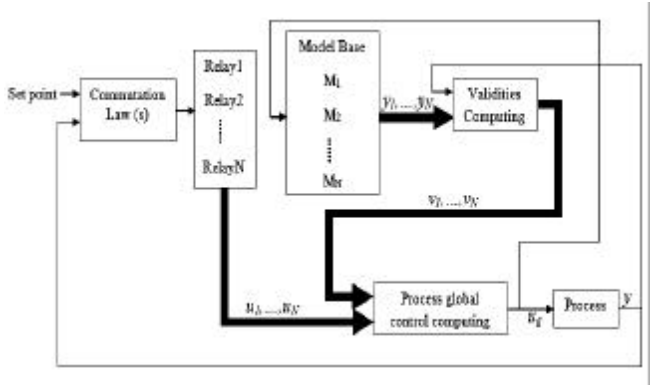

Fig.5—Sliding mode multimodel control (SM-MMC) structure

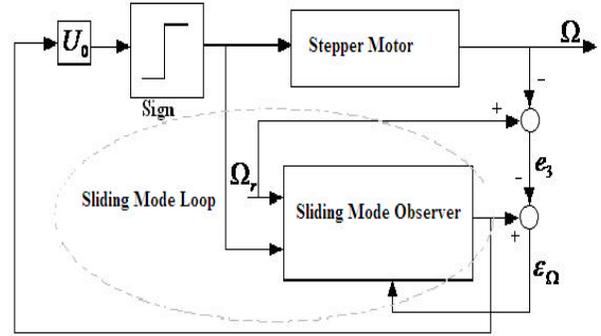

Fig.6—Velocity control based on a sliding mode observer (SMO)

After that, process will converge to the sum of those surfaces weighted by correspondent validities $u_i$ ($S = \sum_i u_i s_i$) and global control will be obtained by adding partials controls $u_i$ ($u_i = u_{ei} + u_{si}$) weighted by adapted validities computed on line ($u_g = \sum_{i=1}^{N} n_i u_i$) with $u_{ei}$, equivalent control relative to each sliding surface. In this way, to improve performances of this control, a PID sliding surface[27-28] was used, and designed as

$$s_i = a_i(x_d - x_r) + b_i \frac{d(x_d - x_r)}{dt} + g_i \int_0^t (x_d - x_r)dt \quad …(11)$$

Then, $\dot{s}_i = a_i \dfrac{d(x_d - x_r)}{dt} + b_i \dfrac{d^2(x_d - x_r)}{dt^2} + g_i(x_d - x_r)$

with $a_i, b_i, g_i > 0$, $i=1,...,N$, $x_d$ the desired state and $x_r$ the real state. To reduce CP, saturation function was used to give

$$u_{si} = l_i sat\left(\frac{s_i}{\Psi_i}\right) \quad …(12)$$

where

$$u_{si} = \begin{cases} l_i sign(s_i) & \text{if } |s_i| \geq \Psi_i \\ l_i\left(\dfrac{s_i}{\Psi_i}\right) & \text{if } |s_i| \leq \Psi_i \end{cases} \quad …(13)$$

with $l_i$ and $\Psi_i > 0$, $\Psi$ defines the boundary layer thickness.

**Sliding Mode Observer (SMO)**

Angular velocities of stepper motors measurements could give unreliable results because of disturbances that influence tachymetry sensor. Therefore, angular velocity $\Omega$ is estimated using SMO instead of sensor. This operation would give better results and will reduce the number of sensors, which are very costly and may have hard maintenance skills. SMO can be used in the state observers design, and has the ability to bring coordinates of estimator error dynamics to zero in finite time. SMOs (Fig. 6) have attractive measurement noise resilience, which is similar to a Kalman filter. CP concerned with sliding mode method can be eliminated by modifying SMO gain, without sacrificing the control robustness and precision qualities[29-34].

Observed position error is given as $e_q = q - \hat{q}$ and angular velocity error as $e_\Omega = \Omega - \hat{\Omega}$, with $\hat{q}$ estimated position and $\hat{\Omega}$ estimated angular velocity. Angular velocity is equal to the position derivative as $\dot{e}_q = \dfrac{dq}{dt} - \dfrac{d\hat{q}}{dt} = \Omega - \hat{\Omega} = e_\Omega$. In this case, Eq. (1) could be represented as

$$\begin{cases} \dfrac{d\hat{\Omega}}{dt} = \dfrac{K}{J}i_q - \dfrac{f_v}{J}\hat{\Omega} - \dfrac{C_r}{J} + l_2 sign(\Omega - \hat{\Omega}), \text{ with } l_1, l_2 > 0 \\ \dfrac{d\hat{q}}{dt} = \hat{\Omega} + l_1 sign(q - \hat{q}) \end{cases} \quad …(14)$$

and

$$\begin{cases} \dot{e}_q = e_\Omega - l_1 sign(e_q) \\ \dot{e}_\Omega = \dfrac{d\Omega}{dt} - \dfrac{d\hat{\Omega}}{dt} = -\dfrac{f_v}{J}e_\Omega \end{cases} \quad …(15)$$

To ascertain state converge to sliding surface, one has to verify Lyaponov[26-29] stability criterion given by quadratic function $V_1 = \frac{1}{2}e_q^2$. In this way, $\dot{V}_1 = e_q(e_\Omega - I_1 sign(e_q))$. Then, $\dot{V}_1 < 0$ if $I_1 > |e_\Omega|_{max}$ and system dynamic is given by $e_\Omega = I_1 sign(e_q)$ when $e_q \to 0$. Consider now a second Lyaponov quadratic function $V_2 = \frac{1}{2}(e_q^2 + Je_\Omega^2)$. As $e_q = 0$, Eq. (7) gives $\dot{V}_2 = -f_v e_\Omega^2 - e_\Omega(I_2 sign(e_\Omega) - C_r)$, then $\dot{V}_2 < 0$ if $I_2 > |C_r|_{max}$.

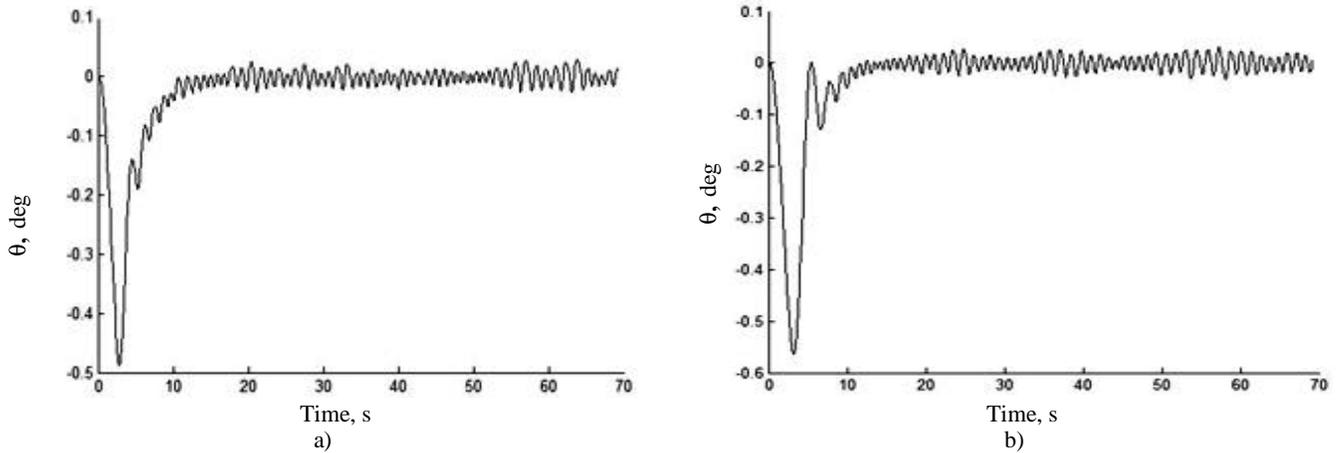

Fig.7—Inclination angle using sliding mode control (SMC) of: a) azimuth motor; and b) altitude motor

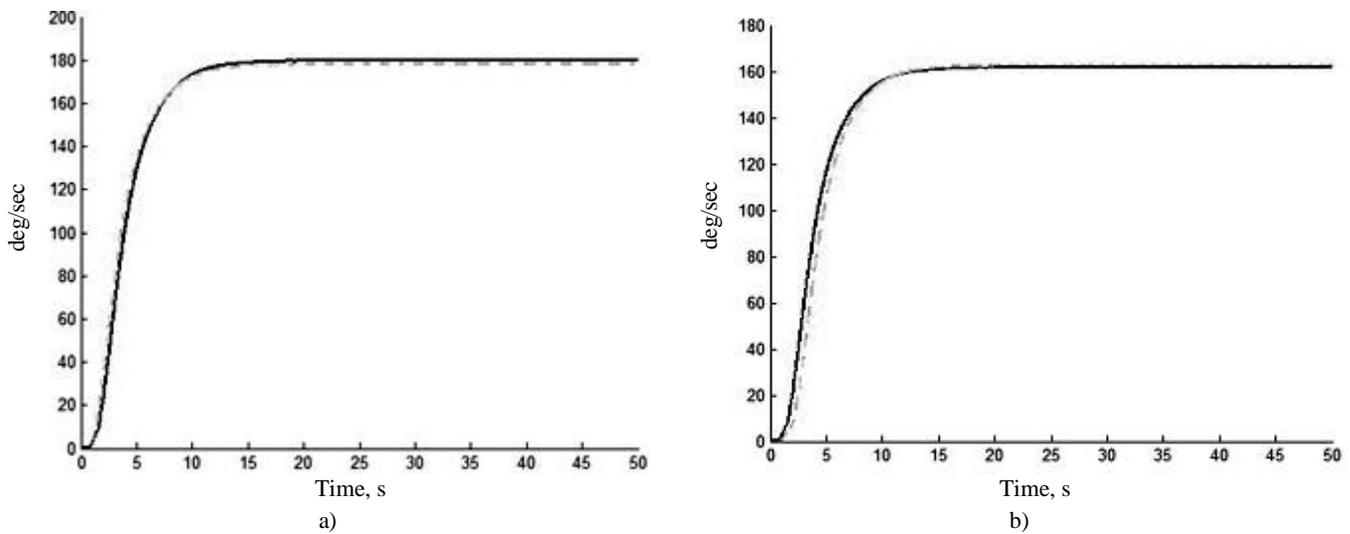

Fig. 8—Real and observed angular velocity of: a) azimuth motor; and b) altitude motor

### Results and Discussion

Experimental results were accomplished for a second order SMC of Eq. (2) applied on the system (Fig. 3) connected to a computer through a serial cable RS232. This required system control evolution, real and observed angular velocities, $\Omega_1$ and $\Omega_2$, and azimuth and altitude inclinations, $\theta_1$ and $\theta_2$. Sliding surface parameters used were $n$ =0.135, $m_1$ =1.2 and $m_2$ =0.355. Stepper motor characteristics were as follows: inertia moment J, 3.0145 $10^{-4}$kg.m$^2$; mechanical torque C, 0.780 Nm; torque constant K, 0.433 Nm/A; armature resistance R, 3.15 Ohm; armature inductance L, 8.15 mH; and friction forces $f_v$, 0.0172 Nms/rad. Experimental test was considered for time exceeding 100 s.

Using second order SMC, it was observed (Fig. 7) that steady state is reached in a short time (~15 s). Also, for the first test, $\theta_1$=0.48° and $\theta_2$=0.57°; thus azimuth inclination $\theta_1$ is very close to altitude inclination $\theta_2$. This fact is related to elliptic movement nature of sun. However, presence of CP (~0.01°) was noticed. Real and observed angular velocities (Fig. 8) of two axis tracker are very close; $\Omega_1$=180°/s $\approx \hat{\Omega}_1$ and $\Omega_2$=160°/s $\approx \hat{\Omega}_2$. This excellent result is due to high estimation quality of SMO. Moreover, SMC shows its robustness (Fig. 7)

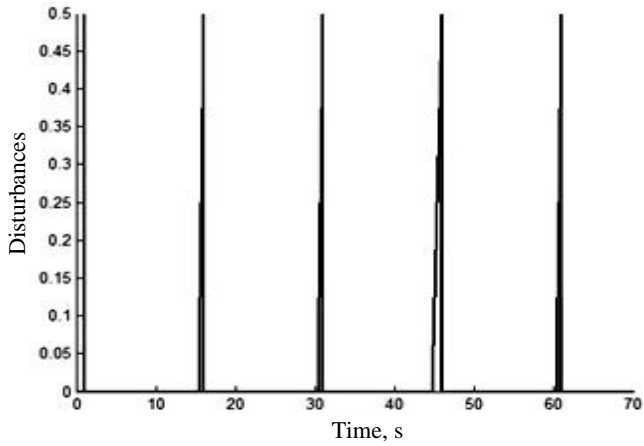

Fig.9—Disturbance signal

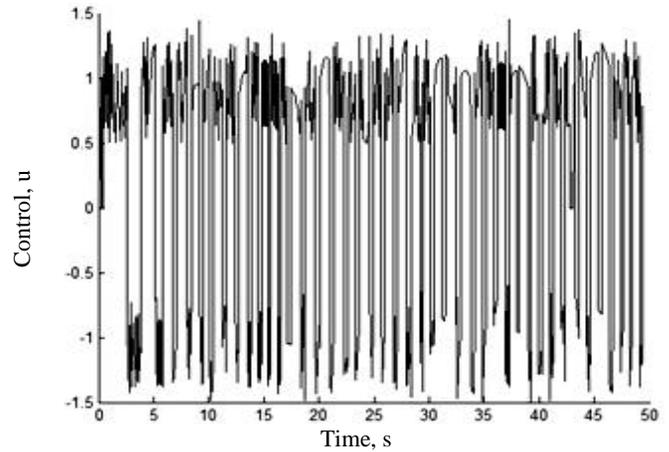

Fig.10—SMC evolution

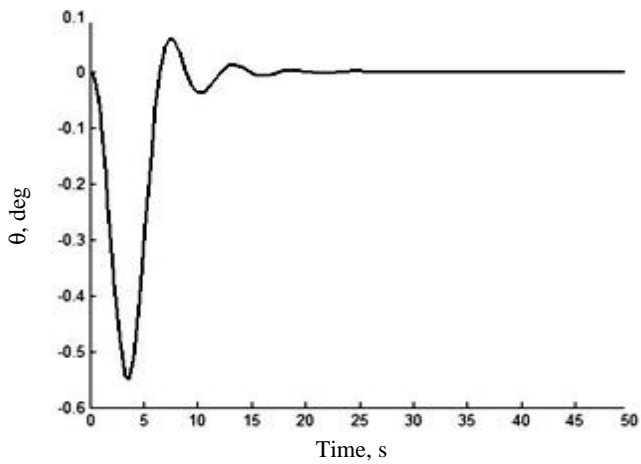

a)

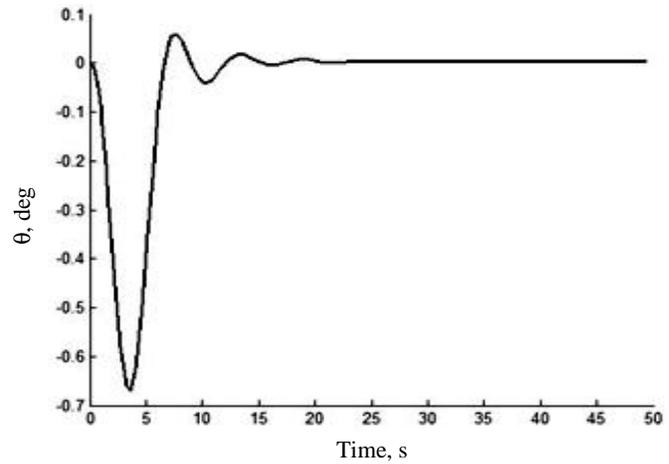

b)

Fig.11—Inclination angle using SM-MMC of: a) azimuth motor; and b) altitude motor

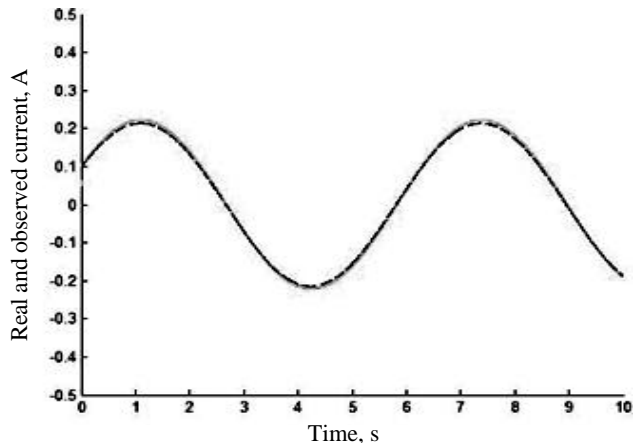

Fig. 12—Real and observed current of the control

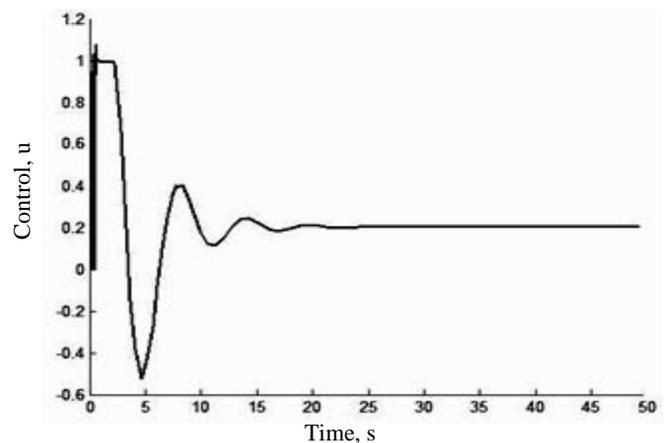

Fig.13—SM-MMC evolution

when a periodic disturbance signal is injected (Fig. 9) with a signal period T of 15 s. System does not deviate from trajectory and preserve its equilibrium position. However, control evolution (Fig. 10), with no high level but with sharp commutations frequency, does not give adequate operating conditions for actuators (both azimuth and altitude motors). As a solution to these problems (chattering and control oscillations), system was simulated by PID SM-MMC approach, resulting in CP elimination (Fig. 11). Also, control evolution (Figs 12 & 13) of the system had less commutation frequency and reached steady state in a very short time, which preserved robustness of the system and gave good conditions for actuators.

## Conclusions

In this study a SMO has been considered to evaluate angular velocities of stepper motors. Experimental results showed effectiveness of SMC in tracking process and its robustness unless disturbances and high estimation quality of SMO. PID SM-MMC was designed and process simulated. This new approach showed excellent result in term of CP elimination and control stabilization with preserving control robustness.